\documentclass{aastex62}
\usepackage{comment}
\usepackage{amsmath}
\usepackage{color}
\usepackage{natbib}
\usepackage{amssymb}
\usepackage{amsfonts}
\usepackage{stackengine}
\newcommand\stackequal[2]{%
	\mathrel{\stackunder[2pt]{\stackon[4pt]{=}{$\scriptscriptstyle#1$}}{%
			$\scriptscriptstyle#2$}}}

\newcommand{\beq}{\begin{equation}}
\newcommand{\eeq}{\end{equation}}
\newcommand{\g}{\gamma}
\def\ba{\begin{eqnarray}}
\def\ea{\end{eqnarray}}
\newcommand{\rc}{\rho_7}
\newcommand{\nc}{\nu_{9}}

\newcommand{\et}{E_{\rm tot,50}}
\newcommand{\Rc}{R_{\rm pc}}
\newcommand{\dc}{D_{\rm Gpc}}
\newcommand{\ee}{\epsilon_e}
\newcommand{\eb}{\epsilon_b}
\newcommand{\tc}{t_{d,2}}
\newcommand{\gc}{\gamma_c}
\newcommand{\gm}{\gamma_{\rm max,6}}

\shorttitle{frb counterparts and central engines} \shortauthors{Wang \& Lai}
\begin{document} 
\title{Fast radio burst counterparts and their implications for the central engine}
\author[0000-0002-2662-6912]{Jie-Shuang Wang}
\correspondingauthor{Jie-Shuang Wang}
\email{jiesh.wang@gmail.com}
\affiliation{Tsung-Dao Lee Institute, Shanghai Jiao Tong University, Shanghai 200240, China}
\author{Dong Lai}
\affiliation{Cornell Center for Astrophysics and Planetary Science, Department of Astronomy, Cornell University, Ithaca, NY 14853, USA}
\affiliation{Tsung-Dao Lee Institute, Shanghai Jiao Tong University, Shanghai 200240, China}
\begin{abstract}
While the radiation mechanism of fast radio bursts (FRBs) is unknown, coherent curvature radiation and synchrotron maser are promising candidates. 
We find that both radiation mechanisms work for a neutron star (NS) central engine with $B\gtrsim 10^{12}$\,G, while for the synchrotron maser, the central engine can also be an accreting black hole (BH) with $B\gtrsim 10^{12}$\,G and a white dwarf (WD) with $B\sim 10^8-10^9$\,G. 
We study the electromagnetic counterparts associated with such central engines, i.e., nebulae for repeating FRBs and afterglows for non-repeating FRBs. 
In general, the energy spectrum and flux density of the counterpart depend strongly on its size and total injected energy.
We apply the calculation to the nebula of FRB 121102 and find that the persistent radio counterpart requires the average energy injection rate into the nebula to be between $2.7\times10^{39}~{\rm\,erg/s}$ and $1.5\times10^{44}~{\rm\,erg/s}$, and the minimum injected energy be $6.0\times10^{47}~{\rm\,erg}$ in around $7$ yr. 
Consequently, we find that for FRB 121102 and its nebula: 
(1) WD and accretion BH central engines are disfavored; 
(2) a rotation-powered NS central engine works when $1.2\times10^{12}~{\rm\,G}\lesssim B\lesssim 7.8\times10^{14}~{\rm\,G}$ with initial period $P<180$\,ms, but the radio emission must be more efficient than that in typical giant pulses of radio pulsars; 
and (3) a magnetic-powered NS central engine works when its internal magnetic field $B\gtrsim 10^{16}$\,G. 
We also find that the radio-emitting electrons in the nebula could produce a significant rotation measure (RM), but cannot account for the entire observed RM of FRB 121102. 
\end{abstract}
\keywords{radio continuum: general -- radiation mechanisms: non-thermal -- stars: magnetars -- relativistic processes}

\section{Introduction}
Fast radio bursts (FRBs) are mysterious radio transients of milliseconds duration. Up to now, more than eighty FRBs have been discovered \citep{Lorimer2007,Keane2012,Thornton2013,Masui2015,Ravi2015,Ravi2016,Champion2016,Petroff2016,Petroff2017,Shannon2018,Boyle2018,Zhang2019,Bannister2019}, among which eleven are found to be repeaters \citep{Spitler2014,Spitler2016,Scholz2016,Chatterjee2017,Law2017,CHIME2019a,CHIMERFRB2019b,KumarASKAP2019}. Due to their large dispersion measures (DM), FRBs are believed to be of extragalactic origin; this is further supported by the direct localization of the host galaxies of FRB 121102 \citep{Chatterjee2017,Marcote2017,Tendulkar2017}, FRB 180924 \citep{Bannister2019}, and FRB 190523 \citep{Ravi2019}.

Many theoretical models of (repeating) FRBs invoke compact stars, especially highly magnetized neutron stars \citep[see the recent reviews by][and references therein]{Platts2018, Cordes2019}. Activities from such compact stars can drive strong outflows, leading to the formation of nebulae or afterglows. Such a nebula has been suggested as an explanation of the persistent radio counterparts associated with FRB 121102 \citep{Chatterjee2017,Dai2016,Murase2016,Metzger2017,Cao2017,Beloborodov2017,Waxman2017,Xiao2017,Margalit2018a,Margalit2018b,YangDai2019}. 

It is of great importance to study the possible electromagnetic counterparts of the FRBs. It can not only help to identify the host galaxies of the FRBs, but also serve as an important tool to study the progenitors and their ambient environment. \cite{Yang2016} studied the FRB-heated nebula and found that there will be a hump in the spectrum. Recently, \cite{Yang2019} found that the optical transient due to inverse Compton scatterings of the prompt radiation of FRBs is generally too weak to be detectable. \cite{Metzger2019} studied the X/gamma-ray transient emissions based on the model of decelerating relativistic blast waves. \cite{Beloborodov2019} found that under certain conditions the blast waves from magnetar flares could lead to optical flashes instead of FRBs. 
In this paper, we will not assume any specific models of FRBs. Instead, we combine the current understanding of the FRB radiation mechanisms and the nebula associated with FRB 121102 to constrain the FRB central engine in a general way. On this basis, we predict the general property of the nebulae and their rotation measure (RM) for repeating FRBs and afterglows for non-repeating FRBs. 

The paper is organized as follows: In Section 2, we briefly review the radiation mechanisms for FRBs and discuss their implications on the central engine. In Section 3, we study the nebula or afterglow powered by the outflow from the same central engine, and we constrain the central engine of FRB 121102 by modeling its nebula in details. The conclusion and discussion are given in Section 4. 

\section{Radiation mechanism and central engines of FRBs}

The radiation of FRBs must be coherent due to their high brightness temperature $T\gtrsim 10^{35}$ K \citep[e.g.][]{Lyutikov2017,Lu2018}. \cite{Lu2018} reviewed a variety of coherent radiation mechanisms, and favoured coherent curvature radiation around magnetic neutron stars \citep[e.g.][and references therein]{Kumar2017,Yang2018}. Recent studies suggest that synchrotron maser may also be possible but with a very low radiation efficiency \citep[e.g.][and references therein]{Plotnikov2019}. We summarize these two radiation mechanisms below, and then place some constraints on the central engine. 
\subsection{Coherent curvature radiation}
Coherent curvature radiation is produced by bunches of coherently moving electrons. The Lorentz factor of the electrons 
emitting GHz radiation in magnetic field lines with a curvature radius $\rho=10^7\rc$ cm is 
\beq
\g\approx110 \rc^{1/3} \nc^{1/3},\label{eq:gammae}
\eeq
where $\nc=\nu/(10^9$ Hz). The radiation power of a single electron is 
\beq
P_e=2\g^4e^2c/(3\rho^2).\label{eq:Pe}
\eeq 
If a coherent bunch contains $N_{e,{\rm coh}}$ net electrons, the total emitted power will be enhanced to 
$P_{\rm coh}=N_{e,{\rm coh}}^2 P_e$. The number of electrons in a bunch is 
\beq
N_{e,{\rm coh}}=n_{\rm coh}V_{\rm B}=1.1\times 10^{21} n_{\rm coh,12}\rc^{2/3}\nc^{-7/3},
\eeq
where $n_{\rm coh}=10^{12}n_{\rm coh,12}$ cm$^{-3}$ is the electron density, and the volume of the bunch is $V_{\rm B}=S_{\rm B}\lambda$ with an area\footnote{The electrons within this area can be causally connected while radiating, see \cite{Lu2018}. The area could be larger, if only the geometric effect of the magnetic field is considered, see \cite{Cordes2016} and \cite{Yang2018}.} $S_{\rm B}=\pi\g^2\lambda^2$, and $\lambda=c/\nu$. Suppose there are $N_{\rm B}=10^5N_{\rm B,5}$ bunches moving toward the observer within the angle of $\g^{-1}$ at the same time, the observed flux density at the luminosity distance $D=\dc$ Gpc is 
\beq
F_{{ \rm FRB},\nu}\approx N_{\rm B} N_{e,{\rm coh}}^2 \g^4 P_e D^{-2}\nu^{-1}=1.3 N_{\rm B,5}n_{\rm coh,12}^2 \rc^2\nc^{-3}D_{\rm Gpc}^{-2}~{\rm Jy}. \label{eq:F_FRB}
\eeq
The required radiation surface area is $S\sim N_{\rm B}S_{\rm B}=3.5\times 10^{12} N_{\rm B,5} \rc^{2/3}\nc^{-4/3}$ cm$^2$. 

In the ``lab'' frame, the radiation formation time of a bunch is 
\beq 
t_{\rm r}\approx \g^2\nu^{-1} \approx 1.2\times 10^{-5}\rc^{2/3}\nc^{-1/3}~{\rm s}.
\eeq
Thus, bunches must form continuously to make a FRB, since the duration of FRBs ($\sim \g^2$ ms in the ``lab'' frame) is much longer. The typical cooling time for the bunch is, 
\beq
t_{\rm c}\approx \g m_e c^2/ (N_{e,{\rm coh}}P_e)=1.3\times 10^{-11} \rc^{1/3} \nc^{4/3}n_{\rm coh,12}^{-1}~ {\rm s}. \label{eq:tc}
\eeq
This is much shorter than the radiation time $t_{\rm r}$.
One possible solution is that there exists an electric field 
continuously accelerating the electron bunches, as suggested by \cite{Kumar2017}. The required local magnetic field for the acceleration is
\beq
B_{\rm acc}\gtrsim E_{\rm acc}\sim N_{e,{\rm coh}}P_e/ec=5.3\times 10^5n_{\rm coh,12}\nc~{\rm G}.
\eeq

Alternatively, a bunch may contain many $e^{\pm}$ pairs $N_p$, yet the net charge is small \citep{Cordes2016}. In this case, the cooling time becomes $t'_{\rm c}=N_p\g m_e c^2/ (N_{e,{\rm coh}}^2P_e)$. To match the radiation formation time $t_{\rm r}$, 
the pair density should satisfy $n_p=N_p/V_{\rm B}=1.0\times 10^{18} n_{\rm coh,12}^2\rc^{1/3}\nc^{-5/3}$. The Goldreich-Julian (GJ) density of a millisecond magnetar is $n_{\rm GJ}=6.90\times 10^{15}B_{14}P_{-3}^{-1}(r/R_{\rm NS})^{-3}$ cm$^{-3}$, where $B=B_{14}10^{14}$ G and $P=P_{-3}10^{-3}$ s. Therefore, a multiplicity $\mathcal{M}=n_p/n_{\rm GJ} =1.5\times 10^2 n_{\rm coh,12}^2\rc^{1/3}P_{-3}(r/R_{\rm NS})^{3}\nc^{-5/3}B_{14}^{-1}$ is required, and this is achievable for a magnetar \citep[see e.g.][and references therein]{Medin2010}.

These moving charged bunches will induce a magnetic field $(B_{\rm ind})$. 
To ensure that this induced field does not significantly change the radiation direction, the local field must satisfy the condition \citep{Lu2018}, 
\beq
B>\g B_{\rm ind}\approx\g4\pi jN_{B}^{1/2}S_{\rm B}^{1/2}/c =1.4\times 10^{12}n_{\rm coh,12} N_{B,5}^{1/2} \rc^{2/3}\nc^{-1/3}~{\rm G}, \label{eq:Bind}
\eeq
where $j=2 n_e e c$. Combining with Eq. \ref{eq:F_FRB}, we find that the coherent curvature radiation can operate for FRB only when the magnetic field in the emission region satisfies\footnote{Note that our result is two orders of magnitude smaller than Eq. 100 in \citep[][hereafter, LK]{Lu2018}. The difference arises from: (1) LK used $\lambda/2\pi$ to indicate the bunch size, while we use $\lambda$; this gives one order of magnitude difference; (2) LK normalized to the isotropic equivalent luminosity $L_{\rm iso,43}$, while we normalize to the flux density $F_{{ \rm FRB},\nu}/{\rm Jy}$, which corresponds to $L_{\rm iso,43}\sim0.01$; this gives another one order of magnitude difference. (3) LK normalized the curvature radius to $10^6$ cm, while we use $10^7$ cm. }
\beq
B\gtrsim1.2\times 10^{12}(F_{{ \rm FRB},\nu}/{\rm Jy})^{1/2} \rc^{-1/3} \nc^{7/6}D_{\rm Gpc}~{\rm G}. \label{eq:B1}
\eeq

\subsection{Synchrotron maser}
\cite{Lyubarsky2014} and \cite{Beloborodov2017} suggested that the synchrotron maser generated in a relativistic shock can also produce FRBs. The required condition is that the pre-shocked outflow must be highly magnetized with a magnetization parameter $\sigma\gtrsim 0.1$. This leads to a rather low radiation efficiency ($f_r\lesssim 10^{-5}$) in the observed band \citep{Iwamoto2017,Iwamoto2018,Plotnikov2019,Metzger2019}, which is consistent with the non-detection of radio emission from the flare of SGR 1806-20 \citep[see e.g.][]{Lyutikov2017}. 
Close to the source region, the magnetic energy is
\beq
E_B={\sigma\over 1+\sigma} {E_{\rm FRB}\over \Gamma f_r} 
\approx {1\over6}B^2R_{\rm source}^3,
\eeq
which leads to 
\beq 
B\approx10^8 \sigma^{1/2} E_{\rm FRB,40}^{1/2}R_{\rm source,9}^{-3/2}(1+\sigma)^{-1/2}\Gamma_2^{-1/2} f_{r,-5}^{-1/2} ~{\rm G}, \label{eq:B2}
\eeq
where $E_{\rm FRB,40}= E_{\rm FRB}/(10^{40}$ erg) and $E_{\rm FRB,40}/\Gamma$ are the observed and intrinsic FRB energies accounting for the Doppler effect from the post-shock frame with the bulk Lorenz factor $\Gamma=10^2\Gamma_2$, $R_{\rm source,9}=R_{\rm source}/(10^9$ cm) is the source size, and $f_{r,-5}=f_r/10^{-5}$. 

\subsection{Possible Central engines}
According to Eqs. (\ref{eq:B1}) and (\ref{eq:B2}), the central engine must be highly magnetized.
Only strongly magnetized white dwarfs (WDs), neutron stars (NSs), or accreting black holes (BHs) are possible. 
We note that for WD central engines, coherent curvature radiation works only when the magnetic field satisfy $B\gtrsim10^{11}$ G, where we take $\rc=10^3$ in Eq. \ref{eq:B1}.
This is much larger than those observed in Milky Way. 
For example, only $\sim 23\%$ of WDs are magnetized in cataclysmic variable systems with $B=10^3-10^9$ G, and only $\sim1\%$ of them have $B\approx 10^8-10^9$ G \citep{Briggs2018}. 
But the synchrotron maser is still feasible for WD (with a typical radius $R_{\rm source,9}\sim1$) central engines. 

For more compact central engines like neutron stars with $R_{\rm source,9}\approx10^{-3}$, Eq. \ref{eq:B2} becomes $B\approx 7.8\times 10^{12} \sigma^{1/2} E_{\rm FRB,40}^{1/2}(1+\sigma)^{-1/2}\Gamma_2^{-1/2} f_{r,-5}^{-1/2} ~{\rm G}$, which is comparable to the constraint by Eq. \ref{eq:B1}. 
For the accreting black holes, the magnetic field near the horizon and the luminosity may be estimated from \citep{Blandford1977,Lee2000b,Lee2000a,Liu2018}
\ba
B_{\rm H}&\approx & 1.3\times 10^{12}\dot{M}^{1/2}M_{\rm BH,1}^{-1}~~{\rm G},\label{B_BH}\\
\dot{E}_{\rm BZ}&\approx & 1.7\times 10^{20}a_*^2M_{\rm BH}^{2}B_{\rm H}^2\approx 2.9\times 10^{46}\dot{M}a_*^2~~{\rm erg/s},\label{Edot_BH}
\ea
where $M_{\rm BH}=10M_{\rm BH,1}M_\odot$ is the black hole mass, $\dot{M}$ is the accretion rate in units of $M_\odot$/yr, and $a_*$ is the dimensionless spin parameter of the black hole. 
Consequently, super-Eddington accretion rate is required to power FRBs, and the lifetime of the FRB is determined by the accretion timescale.
The magnetic field here is calculated from the balance between the ram pressure and the magnetic pressure near the BH horizon, namely, $B_{\rm H}^2/(8\pi)\approx n_{\rm H}m_p c^2\approx \dot{M}c/(4\pi r_{\rm H}^2)$ \citep{Liu2018}, where $n_{\rm H}$ is the number density of the plasma (we assume mainly protons and electrons here) near the horizon, $m_p$ is the proton mass, and $r_{\rm H}$ is the radius of the horizon. 
The corresponding plasma frequency near the horizon is $\nu_{\rm H}=\sqrt{n_{\rm H} e^2/(\pi m_e)}=6.0\times 10^7\dot{M}^{1/2}M_{\rm BH,1}^{-1}~{\rm GHz}$, which is much larger than the typical FRB frequency. 
Therefore, coherent curvature radiation does not work for the accreting BH scenario.
But the synchrotron maser is still viable, since it could be generated at a much larger radius by the jets. 
And it is interesting to note that episodic behaviors in jets have been already suggested in active galactic nuclei and micro-quasars \citep[e.g.][and references therein]{Shende2019}. 

It has been widely suggested that newly born magnetars may be the central engine of FRBs \citep{Popov2013,Lyubarsky2014,Connor2016,Cordes2016,Lyutikov2016,Katz2016,Metzger2017,Beloborodov2017,Katz2017,Kumar2017,Waxman2017,Nicholl2017,Margalit2018a,Margalit2018b,Metzger2019,Cordes2019}. However, burst phenomena can also be produced by old neutron stars via interaction with companion stars or planets \citep[e.g.][]{Mottez2014,Geng2015,Wang2016,Zhang2016,Dai2016,Wang2018}, or `combed' by some energetic flows \citep{Zhang2017}. Our discussion above suggests that, at least in principle, normal NSs with magnetic field $10^{12}$ G can power FRBs for both radiation mechanisms. For the coherent curvature radiation, the magnetic field of the emission site must satisfy Eq. (\ref{eq:B1}).

\section{Electromagnetic counterparts of FRBs}
The coherent curvature radiation and synchrotron maser limit the central engine to be highly magnetized WDs, NSs, or hyper-critical accreting BHs. 
These central engines usually produce highly magnetized outflows. 
Such magnetized outflows will lead to afterglows for non-repeating FRBs and `nebulae' for repeating FRBs. 
We study these two counterparts below, but the fiducial parameter values are chosen for the nebulae.

Suppose an amount of energy $E_{\rm tot}=10^{50}\et$ ergs is 
injected over a time $t_a$, and distributed uniformly in a ``bubble" with radius $R=\Rc$ pc. The energy is mainly carried by leptons (a fraction of $\ee$) and magnetic fields (a fraction of $\eb$) with $\ee+\eb\approx 1$. The magnetic field in the bubble is 
\beq
B=\sqrt{6 \eb E_{\rm tot}/ R^3}=4.5\times10^{-3}\et^{1/2}\eb^{1/2}\Rc^{-3/2}~~{\rm G}. \label{eq:B}
\eeq

The electrons are injected with a power-law spectrum $dn_e/d\gamma_e=K\gamma_e^{-p}$ in $\gamma_{\rm min} \le\gamma_e\le \gamma_{\rm max}$, 
where $\gamma_{\rm min}$ and $\gamma_{\rm max}$ are the minimum and maximum Lorentz factor, respectively. The power-law index is assumed to be $1<p<2$, as found in the radio counterpart of FRB 121102 \citep{Chatterjee2017}. Such hard electron spectra are also observed in some pulsar wind nebulae \citep[e.g.][]{Chevalier2005} and `magnetar' wind nebula \citep{Granot2017}.
It is noteworthy that simulations of magnetic reconnection in the magnetized plasma $\sigma>10$ also produce electron spectra with $p\approx 1.5$ \citep[see][and reference therein]{Law2019}. The normalization factor $K$ can be obtained by $3 \ee E_{\rm tot}/4\pi R^3 =\int \gamma m_e c^2 dn_e $, which gives
\ba
K \approx \frac{3(2-p)\epsilon_e E_{\rm tot}}{4\pi R^3 m_ec^2\gamma_{\rm max}^{2-p}} \stackequal{}{p=1.5}5.0\times10^{-4}\ee\et\gm^{-1/2}\Rc^{-3}~{\rm cm}^{-3}\label{Kvalue}, 
\ea
where $\gamma_{\rm max}=10^6\gm$.
The total number of electrons is $N_e=4\pi R^3n_e/3$, with the electron density given by 
\beq 
n_e=\int_{\gamma_{\rm min}}^{\gamma_{\rm max}} dn_e=K(p-1)^{-1}\gamma_{\rm min}^{1-p}. \label{eq:neNEB}
\eeq

Synchrotron radiation cools the electrons. In a dynamical time of $t_d=10^2t_{d,2}$ years, electrons with 
Lorentz factor larger than the critical value $\gc$ will significantly lose their energy \citep{Sari1998}, where 
\beq
\gc=6\pi m_e c/(\sigma_TB^2t_d)=1.2\times10^4\Rc^3\et^{-1}\tc^{-1}\eb^{-1}.
\eeq
For the nebula, the dynamical time equals to the age of the central engine, $t_d=t_a$. 
The typical frequency for synchrotron radiation is 
\beq
\nu(\g)=\gamma^2{eB/(2\pi m_ec)}. \label{nusyn}
\eeq
The corresponding critical synchrotron radiation frequency is 
\beq
\nu_c
=1.8\times10^3\Rc^{9/2}\et^{-3/2}\tc^{-2}\eb^{-3/2}~~{\rm GHz}.\label{nuc}
\eeq
Due to the synchrotron cooling, the steady-state electron spectrum becomes \citep{Sari1998}
\begin{equation}
\frac{dn_e}{d\gamma_e}=\left \{
\begin{array}{ll}
K\gamma_e^{-p}, & \gamma_{\rm min}\le\gamma_e<\gamma_c,\\
K\gamma_c\gamma_e^{-(p+1)}, & \gamma_c\le\gamma_e\le \gamma_{\rm max}.
\end{array}
\right.\label{dne}
\end{equation}
The corresponding synchrotron spectrum is characterized by three break frequencies: the critical frequency ($\nu_c$), the synchrotron self-absorption frequency $(\nu_a)$, and the typical frequency produced by the electrons with the minimum Lorentz factor ($\nu_m$). The later is giving by 
\beq
\nu_m
=1.2\times10^{-5}\gamma_{\rm min}^{2}\et^{1/2}\eb^{1/2}\Rc^{-3/2}~{\rm GHz}. \label{numin}
\eeq
The synchrotron self-absorption frequency is determined by setting the synchrotron optical depth to unity, $\tau=k_{\nu}R=1$, 
where $k_{\nu}$ is the synchrotron self-absorption coefficient. For electrons with a power-law distribution, $\nu_a$ is giving by Eq. 6.53 of \cite{Rybicki1986} or Eqs. (A8, A9) of \cite{Wu2003}, 
\beq
\nu_a=\left \{
\begin{array}{ll}
	\left(\frac{c_2eKR_p}{B}\right)^{2/(p+4)}\frac{eB}{2\pi m_ec} \stackequal{}{p=1.5} 0.03\et^{15/22}\eb^{7/22}\ee^{4/11} \Rc^{-37/22}\gm^{-2/11}~{\rm GHz},~ &\nu_a>\nu_m,\\
	\left(\frac{c_1eKR_p}{B}\right)^{3/5}\gamma_{\rm min}^{-3(p+4)/5}\nu_m\stackequal{}{p=1.5}5.8 \et^{4/5} \eb^{1/5} \ee^{3/5}\Rc^{-9/5}\gm^{-3/10} \gamma_{\rm min}^{-13/10} ~{\rm GHz},~& \nu_a<\nu_m, 
\end{array}\label{eq:nu_a}
\right.
\eeq
Then the overall synchrotron radiation spectrum is \citep{Meszaros1997,Sari1998}
\begin{equation}
F_\nu= \left \{
\begin{array}{lll}
F_{\nu,\rm max}(\nu/\nu_m)^{-(p-1)/2}, & {\rm Max}[\nu_a,\nu_m]<\nu<\nu_c,\\
F_{\nu,\rm max}(\nu_c/\nu_m)^{-(p-1)/2}(\nu/\nu_c)^{-p/2}, & \nu\ge\nu_c.
\end{array}
\right.\label{Fnu}
\end{equation}
The peak flux density at a luminosity distance $D=\dc$ Gpc is 
\ba
\begin{aligned}
F_{\nu,\rm max} =& \frac{N_e}{4\pi D_L^2}\frac{m_ec^2\sigma_T}{3e}B\\
\stackequal{}{p=1.5} & 1.74\times 10^2\et^{3/2}\ee\eb^{1/2}\dc^{-2}\Rc^{-3/2}\gamma_{\rm min}^{-1/2}\gm^{-1/2} ~~\mu {\rm Jy}.\label{Fnumax}
\end{aligned}
\ea

\subsection{Constraints on FRB 121102}
For the persistent radio source associated with FRB 121102 \citep{Chatterjee2017}, 
the spectral index is $\sim-0.25$ for frequencies below $\nu_c$, and $\sim-0.75$ above $\nu_c$. 
This leads to an electron spectral index of $p\approx1.5$. 
The redshift is $z_s=0.193$ \citep{Tendulkar2017}, so the luminosity distance to the observer is $D=0.972$ Gpc. 
The size of this persistent source is found to be $<0.7$ pc \citep{Chatterjee2017,Marcote2017}, 
while FRB 121102 has been repeating for at least 7 years. 
Thus 
\beq
\tc>0.07,~\&~\Rc<0.7.\label{eq:size_age}
\eeq
The critical frequency is $\nu_c\approx10(1+z_s)$ GHz, and the flux density at the critical frequency is $F_{\nu}(\nu_c) \approx 166~\mu$Jy \citep{Chatterjee2017}. 
Combining with Eqs. (\ref{nuc}), (\ref{Fnu}), and (\ref{Fnumax}), we find 
\ba
\eb\approx28.6\Rc^3\et^{-1}\tc^{-4/3},\label{eq:eb_con}\\
\ee\approx3.4\tc^{5/6}\gm^{1/2}\et^{-1}.\label{eq:ee_con}
\ea
The condition $\eb+\ee\approx 1$ translates to 
\beq
\et\approx28.6\Rc^3\tc^{-4/3}+3.4\tc^{5/6}\gm^{1/2}\label{et}.
\eeq

The absence of spectral break in the frequency range of $1.6-10$ GHz requires that $\nu_m<1.6(1+z_s)$. Substituted with Eqs. (\ref{eq:eb_con})-(\ref{eq:ee_con}), it leads to
\beq
\gamma_{\rm min}<1.7\times 10^2\tc^{1/3}\label{gmin1}. 
\eeq
The detection of FRB 121102 in the 600 MHz band \citep{Josephy2019} leads to $\nu_a<0.6(1+z_s)$ GHz. 
Combining it with Eqs. (\ref{eq:eb_con})-(\ref{eq:ee_con}),  we obtain 
\ba
\gamma_{\rm min}&>&14.9 \tc^{7/39}\Rc^{-12/13},~\&~ \gamma_{\rm min}>48.0\tc^{3/11}\Rc^{-4/11},~~{\rm for~\nu_a <\nu_m};\nonumber \\ 
\Rc&>&0.12\tc^{-1/6},~\&~\gamma_{\rm min}<46.8\tc^{3/11}\Rc^{-4/11},~~{\rm for~\nu_a>\nu_m}.\label{gmin2}
\ea
Combining with the Eq. (\ref{eq:size_age}), we find that for the $\nu_a <\nu_m$ case, it translates to the constraint $\gamma_{\rm min}>26.4$;
For the $\nu_a >\nu_m$ case, the requirement of $\tc>2.5\times 10^{-5}$ from $0.7>\Rc>0.12\tc^{-1/6}$ is naturally satisfied. 


The maximum frequency must be larger than $22$ GHz based on observation, which gives a lower limit of $\gamma_{\rm max}$ (see below) according to Eq. (\ref{nusyn}). The upper limit of $\gamma_{\rm max}$ can be obtained from the balance between the acceleration and the synchrotron cooling, i.e. $P_{\rm acc}\approx eBc\simeq P_{\rm syn}\approx0.67e^4B^2\gamma_{\rm max}^2m_e^{-2}c^{-3}$.
Combining with Eqs. (\ref{eq:B}) and (\ref{eq:eb_con}), we obtain
\beq
6.2\times10^{-4}\tc^{1/3}<\gm<6.12\times 10^2\tc^{1/3}.\label{gmax}
\eeq

Eqs. (\ref{eq:size_age}), (\ref{et}), and (\ref{gmax}) imply $\et/\tc>3.4\tc^{-1/6}\gm^{1/2}=0.086$ and $\et/\tc<4.9\times 10^3$. 
Therefore, the average energy injection rate into the nebula must satisfy
\beq
2.7\times10^{39}~{\rm erg/s}<\dot{E}<1.5\times10^{44}~{\rm erg/s}. \label{edot_con}
\eeq
A minimum energy budget can be obtained from Eqs. (\ref{eq:size_age}) and (\ref{edot_con}), 
\beq
E_{\rm tot}=\dot{E}t_d>6.0\times10^{47}~{\rm erg}. \label{e_con}
\eeq
Eqs. (\ref{edot_con}) and (\ref{e_con}) are one of the key results of this paper. 
They provide useful constraints on the central engine of FRB 121102 and its nebula. 
We consider the following cases:

(i) For a magnetized WD with $B=10^9$ G, the total magnetic energy is $2\times 10^{44}$ erg, which is much smaller than the required minimum energy. 
Therefore, the magnetized WD central engine can be ruled out for FRB 121102. 

(ii) For a NS central engine driven by magnetic energy, the available internal magnetic energy is $2\times 10^{49}B_{*,16}^2R_{\rm NS,6}^3$ erg, where $B_*=B_{*,16}10^{16}$ G is the internal magnetic field and $R_{\rm NS}=10^6R_{\rm NS,6}$ cm is the radius. 
The ambipolar diffusion timescale is around $374 B_{*,16}^{-1.2}$ years \citep{Beloborodov2016}. 
The corresponding average luminosity is then $1.4\times 10^{39}B_{*,16}^{3.2}R_{\rm NS,6}^3$ erg/s. 
Comparing with Eq. (\ref{edot_con}), we find 
\beq
B_{*,16}\gtrsim1.\label{magnetar_con}
\eeq

(iii) For a NS central engine driven by rotational energy, the total available energy is $E_{\rm rot}=0.5 I (2\pi/P_0)^2=2.0\times10^{52} I_{45}P_{0,-3}^{-2}$ erg, where $I=10^{45}I_{45}$ is the moment of inertia, and $P_0=P_{0,-3}10^{-3}$ s is the initial period. 
Combining with Eq. (\ref{e_con}), we obtain $P_{0,-3}<180 I_{45}^{1/2}$. 
It should be noted that the NS initial period could be much smaller than this value, since the new-born NS could dissipate its energy mainly into supernova ejecta at the very beginning.
The NS spin-down luminosity is 
$
L_{\rm sd}={E_{\rm rot}\over\tau_{\rm sd}}{1\over(1+t/\tau_{\rm sd})^2}=4.8\times10^{42}B_{12}^{2}R_{\rm NS,6}^6P_{0,-3}^{-4}{1\over(1+t/\tau_{\rm sd})^2}~{\rm erg/s},
$
where $\tau_{\rm sd}=3c^3IP_0^2/(2\pi^2B_{12}^{2}R_{\rm NS,6}^6)=130 I_{45}P_{0,-3}^2B_{12}^{-2}R_{\rm NS,6}^{-6}$ yr is the spin-down time, and $B=10^{12}B_{12}$ G is the surface magnetic field. 
As the central engine of FRB 121102 keeps active for 7 years, and the burst's peak flux density does not decrease during this period, we require the spin-down time to satisfy $\tau_{\rm sd}>$ 7 yrs, which implies $B_{12}<4.3 I_{45}^{1/2}R_{\rm NS,6}^{-3} P_{0,-3}$. 
On the other hand Eq. (\ref{edot_con}) requires $0.024P_{0,-3}^2R_{\rm NS,6}^{-3}<B_{12}<5.6 P_{0,-3}^2R_{\rm NS,6}^{-3}$. 
Combining with Eq. (\ref{eq:B1}), we obtain 
\beq
{\rm Max}[0.024R_{\rm NS,6}^{-3}P_{0,-3}^2,1.2(F_{{ \rm FRB},\nu}/{\rm Jy})^{1/2} \rc^{-1/3} \nc^{7/6}D_{\rm Gpc}]<B_{12}<{\rm Min}[5.6 R_{\rm NS,6}^{-3}P_{0,-3}^2,4.3 I_{45}^{1/2}R_{\rm NS,6}^{-3} P_{0,-3}]. 
\eeq
Using $P_{0,-3}<180 I_{45}^{1/2}$, we obtain 
\beq
1.2(F_{{ \rm FRB},\nu}/{\rm Jy})^{1/2} \rc^{-1/3} \nc^{7/6}D_{\rm Gpc}<B_{12}<780 I_{45}R_{\rm NS,6}^{-3}. \label{B_NS_con}
\eeq
We note that giant pulse from radio pulsars is powered by such a central engine. 
However, the maximum radiation efficiency (the ratio between the luminosity of the radio pulse and the spin-down luminosity) of the giant pulse is smaller than $10^{-2}$ \citep{Cordes2016}. 
Therefore, for a `giant-pulse-like' FRB, the radio luminosity would be $<10^{42}$ erg/s according to Eq. (\ref{edot_con}), which is smaller than the luminosity of some strong bursts ($\sim 10^{43}$ erg/s) of FRB 121102. 
Consequently, a higher radiation efficiency is required to power FRBs by 
`giant-pulse-like' mechanisms for such a central engine. 

(iv) For an accreting BH central engine, Eqs. (\ref{eq:B1}) and (\ref{B_BH}) require $\dot{M}>M_{\rm BH,1}^{2}$. 
Combining this with Eqs. (\ref{Edot_BH}) and (\ref{edot_con}), we obtain  
\beq
3.1\times10^{-4}\dot{M}^{-1/2}<a_*<7.2\times10^{-2}M_{\rm BH,1}^{-1}. 
\eeq
However, we note that for such a central engine, a high accretion rate must be maintained for at least 7 years. 
In the collapsar model, such a high accretion rate can only last for a few days, since it drops as $t^{-5/3}$ \citep[e.g.][]{MacFadyen2001,Zhang2008}. 
Therefore, it is unlikely that a repeating FRB elder than a month can be powered by such a central engine.

\subsection{DM and RM from the nebula associated with FRB 121102}
The electrons in the nebula can also produce RM, which can be expressed as 
\beq
\text{RM}\approx {e^3\over 2\pi m_e^2 c^4 \gamma_{\rm min}^2} n_e B_{\parallel}R.
\eeq
We assume $B_\perp\approx B_\parallel\approx B$ and the coherent length of the magnetic field is of order the size of the nebula. 
The RM of the nebula associated with FRB 121102 is then obtained by substituting Eqs. (\ref{eq:B}), (\ref{eq:neNEB}), (\ref{eq:eb_con}) and (\ref{eq:ee_con}) into the above equation:  
\beq
\text{RM}\approx 67.1 \tc^{1/6}\Rc^{-2}\gamma_{\rm min}^{-5/2} ~{\rm rad~m}^{-2}.\label{Eq:RM}
\eeq
Combining with Eq. (\ref{eq:size_age}), we obtain ${\rm RM}\gtrsim 87.9\gamma_{\rm min}^{-5/2}$ rad m$^{-2}$. 
According to Eq. (\ref{gmin2}), the RM will be highly suppressed for the $\nu_a<\nu_m$ case, as it requires $\gamma_{\rm min}>26.4$. 
We thus consider the $\nu_a>\nu_m$ case, which allows $\gamma_{\rm min}\sim2$, but requires $\Rc>0.12\tc^{-1/6}$. 
Substituting it into Eq. (\ref{Eq:RM}), we obtain $\text{RM}<4.7\times10^3\tc^{1/2}\gamma_{\rm min}^{-5/2}~{\rm rad~m}^{-2}$.
For a dynamical timescale of $\tc\lesssim10$, the RM from the radio-emitting electrons in the nebula  is an order of magnitude smaller than the observed value of FRB 121102, $\text{RM}=1.03\times10^5~{\rm rad~m}^{-2}$  \citep{Michilli2018}. 


The dispersion measure (DM) of the nebula reads 
\beq
\text{DM}\approx n_e R\gamma_{\rm min}^{-1}\approx 3.4\times10^{-3}\tc^{5/6} \Rc^{-2} \gamma_{\rm min}^{-3/2} ~{\rm pc~cm}^{-3}. 
\eeq 
For FRB 121102, Eq. (\ref{eq:size_age}) leads to $\text{DM}>7.6\times 10^{-4} \gamma_{\rm min}^{-3/2}~{\rm pc~cm}^{-3}$. 
The DM will also be highly suppressed for the $\nu_a<\nu_m$ case; while for the $\nu_a>\nu_m$ case, we obtain $\text{DM}<0.2\tc^{7/6} \gamma_{\rm min}^{-3/2}~{\rm pc~cm}^{-3}$. 

Comparing with the observed  $\text{DM}\approx 559~{\rm pc~cm}^{-3}$, the DM and its variation from the expanding nebula will be negligible ($<0.1\%$) when $\tc<1$.

\subsection{General features on the nebulae and afterglow \label{nebula}}
The general spectral feature for the nebulae and afterglow is described by three 
frequencies: $\nu_c$ (Eq. \ref{nuc}), $\nu_m$ (Eq. \ref{numin}), and $\nu_a$ (Eq. \ref{eq:nu_a}). 
Here we take the electron spectral index to be $p=1.5$. 
The spectral flux density is described by Eqs. (\ref{Fnu}) and (\ref{Fnumax}). 

\begin{figure}[h]
	\begin{center}
	\includegraphics[scale=0.8]{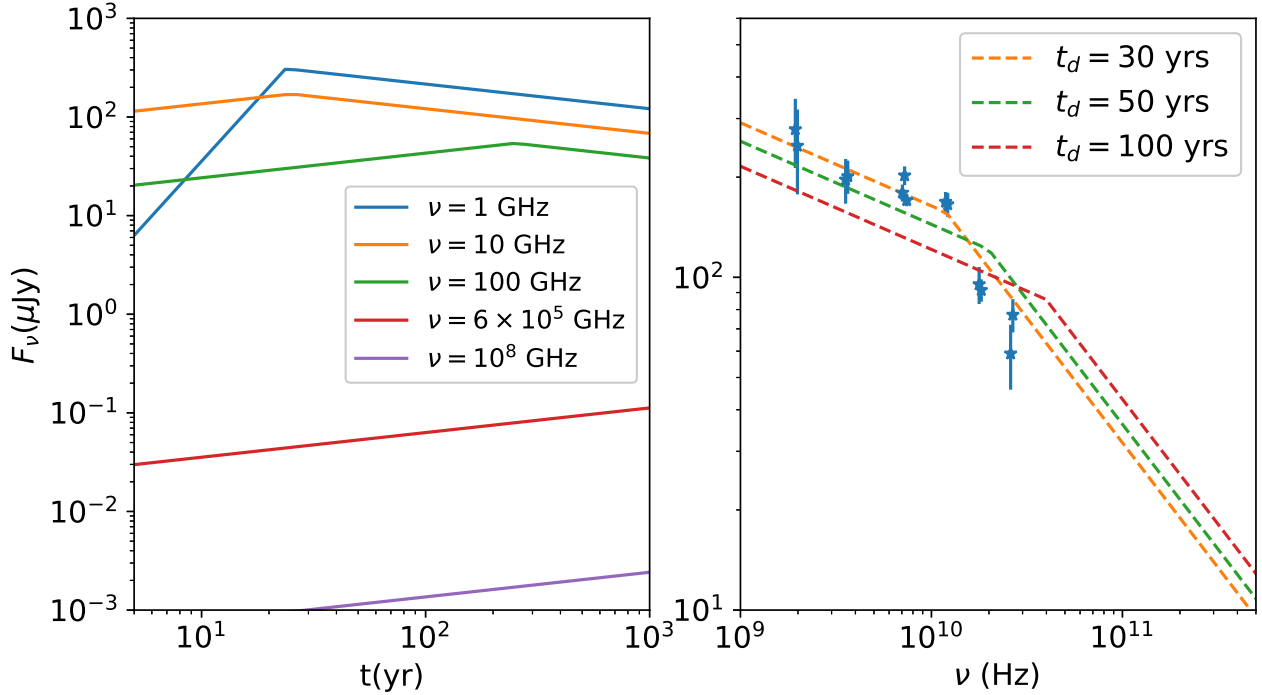}
	\caption{The radio, optical, and X-ray counterparts of a nebula are shown with different colors. The parameters are fixed at $ \dot{E}=10^{41}~{\rm erg/s},~\eb=0.26, \dc=1,~\gamma_{\rm min}=2,~\gm=0.3,~ R=v_0 t$ with $v_0=4.0\times10^8$ cm/s. The left panel shows the time evolution of the flux density in different bands. The right panel shows the spectral evolution at different dynamical time $t_d$ of the nebula. The stars with error bars are the VLA spectrum in different observational time \citep[see the extended data figure 2 of][]{Chatterjee2017}. 
	\label{fig1}}
	\end{center}
\end{figure}

For repeating FRBs, a nebula can be produced. 
An example of the radio, optical and X-ray flux densities of a FRB nebula, under the assumption of $E_{\rm tot}=\dot{E}t_d$ and $R= v_0 t_d$ with the nebula expansion speed $v_0=4\times10^8$ cm/s, is shown in the left panel of Figure \ref{fig1}. 
In the right panel, we compare the spectrum to the observation of FRB 121102 \citep{Chatterjee2017}. 
Note that the parameter fits are not unique. 
The increasing phase of the flux density at $10,~100,~6\times10^5$, and $10^8$ GHz arises because the critical frequency increases with time, while the reason for the increasing phase of the flux density at $1$ GHz is that the absorption frequency decreases with time, so that it becomes transparent for $1$ GHz at around few tens of years. 

For repeating FRBs, the RM and DM produced by the nebula are given by,
\ba
\text{RM}&=&3.6\et^{3/2}\eb^{1/2}\ee \Rc^{-7/2}\gm^{-1/2} \gamma_{\rm min}^{-5/2}~~{\rm rad~m}^{-2},\label{eq:nebulaRM} \\\nonumber
&=&
0.02F_{\nu,\rm max}\dc^{2}\Rc^{-2}\gamma_{\rm min}^{-2},\\
\text{DM}&=&9.9\times10^{-4} \et\ee \Rc^{-2}\gm^{-1/2} \gamma_{\rm min}^{-3/2}~~{\rm pc~cm}^{-3}. 
\ea
The scaling law is $\text{RM}\propto F_{\nu,\rm max}R_{\rm pc,-2}^{-2}$. 

In certain parameter regime, the nebula will be peaked in the optical band. 
For example, when $\nu_c \approx 6\times 10^{14}$ Hz, the peak flux density is 
\beq
F_{\nu,\rm max}\approx 25.2\et\ee \dc^{-2}\tc^{-2/3}\gamma_{\rm min}^{-1/2}\gm^{-1/2}~~\mu {\rm Jy}.
\eeq

\begin{figure}[h]
	\begin{center}
		\includegraphics[scale=0.8]{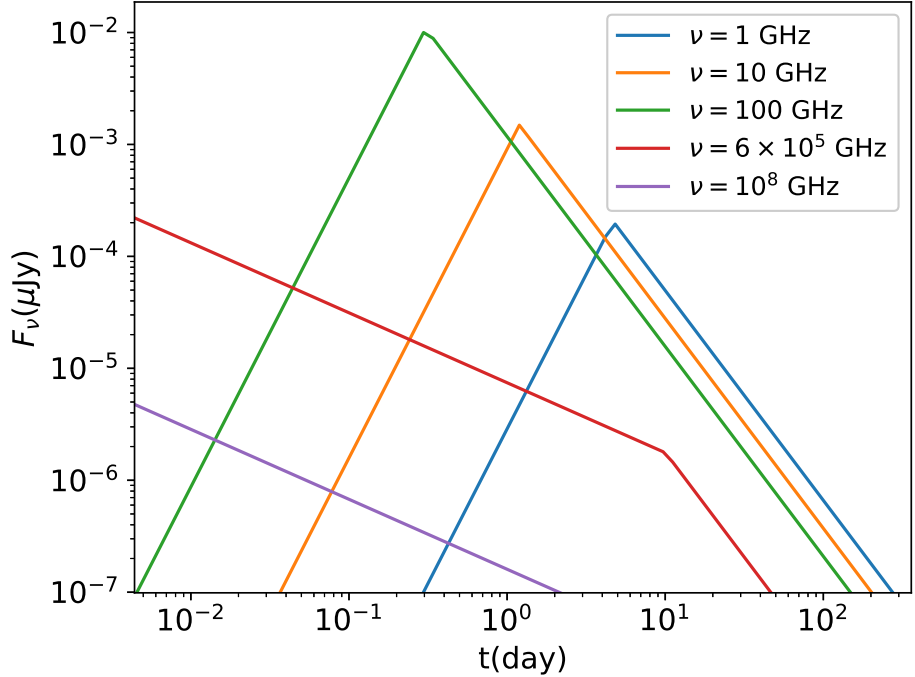}
		\caption{The radio, optical, and X-ray counterparts of a non-repeating FRB afterglow are shown with different colors. The fiducial parameters are fixed at $ E_{\rm tot}=10^{43}~{\rm erg},~\eb=0.26,~\dc=1,~\gamma_{\rm min}=2,~\gm=0.3,~ R=v_0 t$ with $v_0=10^9$ cm/s. \label{fig2}}
	\end{center}
\end{figure}

For a non-repeating FRB, an afterglow can be produced by the outflow. 
In this case, the energy is only injected at the initial time.
An example of the radio, optical and X-ray flux densities, under assumption of $R\propto t_d$, are shown in Figure \ref{fig2}. 
The increasing phase of the flux density in $1$ and $10$ GHz is caused by the decreasing of the absorption frequency. 
This result is consistent with the detection upper limit of the possible counterparts of FRB 180924 \citep{Bannister2019}. 

\section{Conclusion and discussion}
In this paper, we have constrained the central engine of FRBs based on currently available observational data.
We have studied the requirements of the magnetic field for two most commonly discussed FRB radiation mechanisms: Eq. (\ref{eq:B1}) for the coherent curvature radiation and Eq. (\ref{eq:B2}) for the synchrotron maser. 
We find that: (1) NSs with $B\gtrsim 10^{12}$ G can power FRBs for both radiation mechanisms; (2) WDs with $B\sim10^8-10^9$ G and accreting BHs with $B\gtrsim 10^{12}$ G in principle could also be FRB central engine if the radiation mechanism is synchrotron maser.

Since these central engines are very likely to drive highly magnetized outflows associated with FRBs, we have studied the electromagnetic counterparts (nebulae for repeating FRBs, and afterglows for non-repeating FRBs) powered by such magnetized outflows using a simple one-zone model (Sec. \ref{nebula}). 
Applying our general calculations to the nebula of FRB 121102, 
we obtain strong constraints on the energetic properties of the central engine of FRB 121102 and its nebula: the average energy injection rate is limited by Eq. (\ref{edot_con}), and the total energy budget is limited by Eq. (\ref{e_con}). 
These imply that: 
(1) A magnetic WD central engine is ruled out for FRB 121102 because too small amount of energy is available; 
(2) A magnetar central engine powered by pure magnetic energy works for FRB 121102 if the internal magnetic field is larger than $10^{16}$ G (Eq. \ref{magnetar_con}); 
(3) A ``normal'' NS central engine powered by rotational energy also works for FRB 121102 if its magnetic field satisfies $1.2\times10^{12}~{\rm G}\lesssim B\lesssim 7.8\times10^{14}~{\rm G}$ (Eq. \ref{B_NS_con}) with initial period $P<180$ ms, but the radio radiation efficiency must be higher than that for typical giant pulses of regular radio pulsars; 
(4) An accreting BH central engine is disfavored, because it is unclear how to maintain an accretion rate of $M_\odot$/yr for more than 7 years. 
Our study also indicates the radio-emitting electrons in the nebula could produce a significant RM, e.g.  $\text{RM}<4.7\times10^3\tc^{1/2}\gamma_{\rm min}^{-5/2}~{\rm rad~m}^{-2}$ for FRB 121102. But it is unlikely to be responsible for the whole observed RM of FRB 121102, which is $\text{RM}\approx10^5~{\rm rad~m}^{-2}$.

The general properties of the FRB counterparts in our model are summarized in Sec. \ref{nebula}. The peak flux density ($F_{\nu,\rm max}\propto E_{\rm tot}^{3/2}R^{-3/2}$), peak frequency ($\nu_c\propto E_{\rm tot}^{-3/2}R^{9/2}$), and RM ($\propto E_{\rm tot}^{3/2}R^{-7/2}$) are found to be strongly dependent on the size and total injected energy of counterparts. 
This indicates that the observability and spectrum of the counterpart will be significantly affected by the power and repeating rate of the central engine, and the density of the circum-burst medium. 
The later mainly influence the expansion speed of the nebula. 
Recent studies of FRB host galaxies \citep{Bannister2019,Ravi2019} suggest that FRB progenitors can be from an old stellar population, such as magnetars formed after the mergers of compact objects. 
While the nebula from a magnetar born in the compact star merger may expand much faster than the nebula produced after the supernova explosion.
This can lead to lower persistent radiation and RM, but a higher peak frequency,
since Eqs. (\ref{nuc}), (\ref{Fnumax}), and (\ref{eq:nebulaRM}) give $F_{\nu,\rm max}\propto v_0^{-3/2}$, $\nu_c\propto v_0^{9/2}t$, and RM $\propto v_0^{-7/2}t^{-2}$, when assuming $R= v_0 t$ and $E_{\rm tot}=\dot{E} t$ for a nebula.

\acknowledgements
We thank the anonymous referee for valuable comments and suggestions.
We also thank Bing Zhang, Zi-Gao Dai, Wenbin Lu, Ben Margalit, Ruoyu Liu, and Yuanpei Yang for discussions. 
JSW is supported by China Postdoctoral Science Foundation.
\bibliographystyle{aasjournal}
\bibliography{ref}

\begin{thebibliography}{}
\expandafter\ifx\csname natexlab\endcsname\relax\def\natexlab#1{#1}\fi
\providecommand{\url}[1]{\href{#1}{#1}}
\providecommand{\dodoi}[1]{doi:~\href{http://doi.org/#1}{\nolinkurl{#1}}}
\providecommand{\doeprint}[1]{\href{http://ascl.net/#1}{\nolinkurl{http://ascl.net/#1}}}
\providecommand{\doarXiv}[1]{\href{https://arxiv.org/abs/#1}{\nolinkurl{https://arxiv.org/abs/#1}}}

\bibitem[{Bannister {et~al.}(2019)Bannister, Deller, Phillips, Macquart,
  Prochaska, Tejos, Ryder, Sadler, Shannon, Simha, Day, McQuinn, North-Hickey,
  Bhandari, Arcus, Bennert, Burchett, Bouwhuis, Dodson, Ekers, Farah, Flynn,
  James, Kerr, Lenc, Mahony, O{\textquoteright}Meara, Os{\l}owski, Qiu, Treu,
  U, Bateman, Bock, Bolton, Brown, Bunton, Chippendale, Cooray, Cornwell,
  Gupta, Hayman, Kesteven, Koribalski, MacLeod, McClure-Griffiths, Neuhold,
  Norris, Pilawa, Qiao, Reynolds, Roxby, Shimwell, Voronkov, \&
  Wilson}]{Bannister2019}
Bannister, K.~W., Deller, A.~T., Phillips, C., {et~al.} 2019, Science,
  \dodoi{10.1126/science.aaw5903}

\bibitem[{{Beloborodov}(2017)}]{Beloborodov2017}
{Beloborodov}, A.~M. 2017, \apj, 843, L26, \dodoi{10.3847/2041-8213/aa78f3}

\bibitem[{{Beloborodov}(2019)}]{Beloborodov2019}
---. 2019, arXiv e-prints, arXiv:1908.07743.
\newblock \doarXiv{1908.07743}

\bibitem[{{Beloborodov} \& {Li}(2016)}]{Beloborodov2016}
{Beloborodov}, A.~M., \& {Li}, X. 2016, \apj, 833, 261,
  \dodoi{10.3847/1538-4357/833/2/261}

\bibitem[{{Blandford} \& {Znajek}(1977)}]{Blandford1977}
{Blandford}, R.~D., \& {Znajek}, R.~L. 1977, \mnras, 179, 433,
  \dodoi{10.1093/mnras/179.3.433}

\bibitem[{{Briggs} {et~al.}(2018){Briggs}, {Ferrario}, {Tout}, \&
  {Wickramasinghe}}]{Briggs2018}
{Briggs}, G.~P., {Ferrario}, L., {Tout}, C.~A., \& {Wickramasinghe}, D.~T.
  2018, \mnras, 481, 3604, \dodoi{10.1093/mnras/sty2481}

\bibitem[{{Cao} {et~al.}(2017){Cao}, {Yu}, \& {Dai}}]{Cao2017}
{Cao}, X.-F., {Yu}, Y.-W., \& {Dai}, Z.-G. 2017, \apjl, 839, L20,
  \dodoi{10.3847/2041-8213/aa6af2}

\bibitem[{{Champion} {et~al.}(2016){Champion}, {Petroff}, {Kramer}, {Keith},
  {Bailes}, {Barr}, {Bates}, {Bhat}, {Burgay}, {Burke-Spolaor}, {Flynn},
  {Jameson}, {Johnston}, {Ng}, {Levin}, {Possenti}, {Stappers}, {van Straten},
  {Thornton}, {Tiburzi}, \& {Lyne}}]{Champion2016}
{Champion}, D.~J., {Petroff}, E., {Kramer}, M., {et~al.} 2016, \mnras, 460,
  L30, \dodoi{10.1093/mnrasl/slw069}

\bibitem[{{Chatterjee} {et~al.}(2017){Chatterjee}, {Law}, {Wharton},
  {Burke-Spolaor}, {Hessels}, {Bower}, {Cordes}, {Tendulkar}, {Bassa},
  {Demorest}, {Butler}, {Seymour}, {Scholz}, {Abruzzo}, {Bogdanov}, {Kaspi},
  {Keimpema}, {Lazio}, {Marcote}, {McLaughlin}, {Paragi}, {Ransom}, {Rupen},
  {Spitler}, \& {van Langevelde}}]{Chatterjee2017}
{Chatterjee}, S., {Law}, C.~J., {Wharton}, R.~S., {et~al.} 2017, \nat, 541, 58,
  \dodoi{10.1038/nature20797}

\bibitem[{{Chevalier}(2005)}]{Chevalier2005}
{Chevalier}, R.~A. 2005, \apj, 619, 839, \dodoi{10.1086/426584}

\bibitem[{{CHIME/FRB Collaboration}(2018)}]{Boyle2018}
{CHIME/FRB Collaboration}. 2018, The Astronomer's Telegram, 11901, 1

\bibitem[{{CHIME/FRB Collaboration} {et~al.}(2019{\natexlab{a}}){CHIME/FRB
  Collaboration}, {Amiri}, {Bandura}, {Bhardwaj}, {Boubel}, {Boyce}, {Boyle},
  {. Brar}, {Burhanpurkar}, {Cassanelli}, {Chawla}, {Cliche}, {Cubranic},
  {Deng}, {Denman}, {Dobbs}, {Fandino}, {Fonseca}, {Gaensler}, {Gilbert},
  {Gill}, {Giri}, {Good}, {Halpern}, {Hanna}, {Hill}, {Hinshaw}, {H{\"o}fer},
  {Josephy}, {Kaspi}, {Landecker}, {Lang}, {Lin}, {Masui}, {Mckinven},
  {Mena-Parra}, {Merryfield}, {Michilli}, {Milutinovic}, {Moatti}, {Naidu},
  {Newburgh}, {Ng}, {Patel}, {Pen}, {Pinsonneault-Marotte}, {Pleunis},
  {Rafiei-Ravandi}, {Rahman}, {Ransom}, {Renard}, {Scholz}, {Shaw}, {Siegel},
  {Smith}, {Stairs}, {Tendulkar}, {Tretyakov}, {Vanderlinde}, \&
  {Yadav}}]{CHIME2019a}
{CHIME/FRB Collaboration}, {Amiri}, M., {Bandura}, K., {et~al.}
  2019{\natexlab{a}}, \nat, 566, 235, \dodoi{10.1038/s41586-018-0864-x}

\bibitem[{{CHIME/FRB Collaboration} {et~al.}(2019{\natexlab{b}}){CHIME/FRB
  Collaboration}, {:}, {Andersen}, {Band ura}, {Bhardwaj}, {Boubel}, {Boyce},
  {Boyle}, {Brar}, {Cassanelli}, {Chawla}, {Cubranic}, {Deng}, {Dobbs},
  {Fandino}, {Fonseca}, {Gaensler}, {Gilbert}, {Giri}, {Good}, {Halpern},
  {H{\"o}fer}, {Hill}, {Hinshaw}, {Josephy}, {Kaspi}, {Kothes}, {Landecker},
  {Lang}, {Li}, {Lin}, {Masui}, {Mena-Parra}, {Merryfield}, {Mckinven},
  {Michilli}, {Milutinovic}, {Naidu}, {Newburgh}, {Ng}, {Patel}, {Pen},
  {Pinsonneault-Marotte}, {Pleunis}, {Rafiei-Ravandi}, {Rahman}, {Ransom},
  {Renard}, {Scholz}, {Siegel}, {Singh}, {Smith}, {Stairs}, {Tendulkar},
  {Tretyakov}, {Vanderlinde}, {Yadav}, \& {Zwaniga}}]{CHIMERFRB2019b}
{CHIME/FRB Collaboration}, {:}, {Andersen}, B.~C., {et~al.} 2019{\natexlab{b}},
  arXiv e-prints, arXiv:1908.03507.
\newblock \doarXiv{1908.03507}

\bibitem[{{Connor} {et~al.}(2016){Connor}, {Sievers}, \& {Pen}}]{Connor2016}
{Connor}, L., {Sievers}, J., \& {Pen}, U.-L. 2016, \mnras, 458, L19,
  \dodoi{10.1093/mnrasl/slv124}

\bibitem[{{Cordes} \& {Chatterjee}(2019)}]{Cordes2019}
{Cordes}, J.~M., \& {Chatterjee}, S. 2019, arXiv e-prints, arXiv:1906.05878.
\newblock \doarXiv{1906.05878}

\bibitem[{{Cordes} \& {Wasserman}(2016)}]{Cordes2016}
{Cordes}, J.~M., \& {Wasserman}, I. 2016, \mnras, 457, 232,
  \dodoi{10.1093/mnras/stv2948}

\bibitem[{{Dai} {et~al.}(2016){Dai}, {Wang}, {Wu}, \& {Huang}}]{Dai2016}
{Dai}, Z.~G., {Wang}, J.~S., {Wu}, X.~F., \& {Huang}, Y.~F. 2016, \apj, 829,
  27, \dodoi{10.3847/0004-637X/829/1/27}

\bibitem[{{Geng} \& {Huang}(2015)}]{Geng2015}
{Geng}, J.~J., \& {Huang}, Y.~F. 2015, \apj, 809, 24,
  \dodoi{10.1088/0004-637X/809/1/24}

\bibitem[{{Granot} {et~al.}(2017){Granot}, {Gill}, {Younes}, {Gelfand },
  {Harding}, {Kouveliotou}, \& {Baring}}]{Granot2017}
{Granot}, J., {Gill}, R., {Younes}, G., {et~al.} 2017, \mnras, 464, 4895,
  \dodoi{10.1093/mnras/stw2554}

\bibitem[{{Iwamoto} {et~al.}(2017){Iwamoto}, {Amano}, {Hoshino}, \&
  {Matsumoto}}]{Iwamoto2017}
{Iwamoto}, M., {Amano}, T., {Hoshino}, M., \& {Matsumoto}, Y. 2017, \apj, 840,
  52, \dodoi{10.3847/1538-4357/aa6d6f}

\bibitem[{{Iwamoto} {et~al.}(2018){Iwamoto}, {Amano}, {Hoshino}, \&
  {Matsumoto}}]{Iwamoto2018}
---. 2018, \apj, 858, 93, \dodoi{10.3847/1538-4357/aaba7a}

\bibitem[{{Josephy} {et~al.}(2019){Josephy}, {Chawla}, {Fonseca}, {Ng},
  {Patel}, {Pleunis}, {Scholz}, {Andersen}, {Bandura}, \&
  {Bhardwaj}}]{Josephy2019}
{Josephy}, A., {Chawla}, P., {Fonseca}, E., {et~al.} 2019, arXiv e-prints,
  arXiv:1906.11305.
\newblock \doarXiv{1906.11305}

\bibitem[{{Katz}(2016)}]{Katz2016}
{Katz}, J.~I. 2016, \apj, 826, 226, \dodoi{10.3847/0004-637X/826/2/226}

\bibitem[{{Katz}(2017)}]{Katz2017}
---. 2017, \mnras, 469, L39, \dodoi{10.1093/mnrasl/slx052}

\bibitem[{{Keane} {et~al.}(2012){Keane}, {Stappers}, {Kramer}, \&
  {Lyne}}]{Keane2012}
{Keane}, E.~F., {Stappers}, B.~W., {Kramer}, M., \& {Lyne}, A.~G. 2012, \mnras,
  425, L71, \dodoi{10.1111/j.1745-3933.2012.01306.x}

\bibitem[{{Kumar} {et~al.}(2017){Kumar}, {Lu}, \& {Bhattacharya}}]{Kumar2017}
{Kumar}, P., {Lu}, W., \& {Bhattacharya}, M. 2017, \mnras, 468, 2726,
  \dodoi{10.1093/mnras/stx665}

\bibitem[{Kumar {et~al.}(2019)Kumar, Shannon, Osłowski, Qiu, Bhandari, Farah,
  Flynn, Kerr, Lorimer, Macquart, Ng, Phillips, Price, \&
  Spiewak}]{KumarASKAP2019}
Kumar, P., Shannon, R.~M., Osłowski, S., {et~al.} 2019, arXiv:1908.10026

\bibitem[{{Law} {et~al.}(2017){Law}, {Abruzzo}, {Bassa}, {Bower},
  {Burke-Spolaor}, {Butler}, {Cantwell}, {Carey}, {Chatterjee}, {Cordes},
  {Demorest}, {Dowell}, {Fender}, {Gourdji}, {Grainge}, {Hessels}, {Hickish},
  {Kaspi}, {Lazio}, {McLaughlin}, {Michilli}, {Mooley}, {Perrott}, {Ransom},
  {Razavi-Ghods}, {Rupen}, {Scaife}, {Scott}, {Scholz}, {Seymour}, {Spitler},
  {Stovall}, {Tendulkar}, {Titterington}, {Wharton}, \& {Williams}}]{Law2017}
{Law}, C.~J., {Abruzzo}, M.~W., {Bassa}, C.~G., {et~al.} 2017, \apj, 850, 76,
  \dodoi{10.3847/1538-4357/aa9700}

\bibitem[{{Law} {et~al.}(2019){Law}, {Abe}, {Morace}, {Arikawa}, {Sakata},
  {Lee}, {Matsuo}, {Morita}, {Ochiai}, {Liu}, {Yogo}, {Okamoto}, {Golovin},
  {Ehret}, {Ozaki}, {Nakai}, {Sentoku}, {Santos}, {d'Humi{\`e}res}, {Korneev},
  \& {Fujioka}}]{Law2019}
{Law}, K.~F.~F., {Abe}, Y., {Morace}, A., {et~al.} 2019, arXiv e-prints,
  arXiv:1904.02850.
\newblock \doarXiv{1904.02850}

\bibitem[{{Lee} {et~al.}(2000{\natexlab{a}}){Lee}, {Brown}, \&
  {Wijers}}]{Lee2000b}
{Lee}, H.~K., {Brown}, G.~E., \& {Wijers}, R.~A.~M.~J. 2000{\natexlab{a}},
  \apj, 536, 416, \dodoi{10.1086/308937}

\bibitem[{{Lee} {et~al.}(2000{\natexlab{b}}){Lee}, {Wijers}, \&
  {Brown}}]{Lee2000a}
{Lee}, H.~K., {Wijers}, R.~A.~M.~J., \& {Brown}, G.~E. 2000{\natexlab{b}},
  \physrep, 325, 83, \dodoi{10.1016/S0370-1573(99)00084-8}

\bibitem[{{Liu} {et~al.}(2018){Liu}, {Song}, {Zhang}, {Gu}, \&
  {Heger}}]{Liu2018}
{Liu}, T., {Song}, C.-Y., {Zhang}, B., {Gu}, W.-M., \& {Heger}, A. 2018, \apj,
  852, 20, \dodoi{10.3847/1538-4357/aa9e4f}

\bibitem[{{Lorimer} {et~al.}(2007){Lorimer}, {Bailes}, {McLaughlin},
  {Narkevic}, \& {Crawford}}]{Lorimer2007}
{Lorimer}, D.~R., {Bailes}, M., {McLaughlin}, M.~A., {Narkevic}, D.~J., \&
  {Crawford}, F. 2007, Science, 318, 777, \dodoi{10.1126/science.1147532}

\bibitem[{{Lu} \& {Kumar}(2018)}]{Lu2018}
{Lu}, W., \& {Kumar}, P. 2018, \mnras, 477, 2470, \dodoi{10.1093/mnras/sty716}

\bibitem[{{Lyubarsky}(2014)}]{Lyubarsky2014}
{Lyubarsky}, Y. 2014, \mnras, 442, L9, \dodoi{10.1093/mnrasl/slu046}

\bibitem[{{Lyutikov}(2017)}]{Lyutikov2017}
{Lyutikov}, M. 2017, \apj, 838, L13, \dodoi{10.3847/2041-8213/aa62fa}

\bibitem[{{Lyutikov} {et~al.}(2016){Lyutikov}, {Burzawa}, \&
  {Popov}}]{Lyutikov2016}
{Lyutikov}, M., {Burzawa}, L., \& {Popov}, S.~B. 2016, \mnras, 462, 941,
  \dodoi{10.1093/mnras/stw1669}

\bibitem[{{MacFadyen} {et~al.}(2001){MacFadyen}, {Woosley}, \&
  {Heger}}]{MacFadyen2001}
{MacFadyen}, A.~I., {Woosley}, S.~E., \& {Heger}, A. 2001, \apj, 550, 410,
  \dodoi{10.1086/319698}

\bibitem[{{Marcote} {et~al.}(2017){Marcote}, {Paragi}, {Hessels}, {Keimpema},
  {van Langevelde}, {Huang}, {Bassa}, {Bogdanov}, {Bower}, {Burke-Spolaor},
  {Butler}, {Campbell}, {Chatterjee}, {Cordes}, {Demorest}, {Garrett}, {Ghosh},
  {Kaspi}, {Law}, {Lazio}, {McLaughlin}, {Ransom}, {Salter}, {Scholz},
  {Seymour}, {Siemion}, {Spitler}, {Tendulkar}, \& {Wharton}}]{Marcote2017}
{Marcote}, B., {Paragi}, Z., {Hessels}, J.~W.~T., {et~al.} 2017, \apj, 834, L8,
  \dodoi{10.3847/2041-8213/834/2/L8}

\bibitem[{{Margalit} \& {Metzger}(2018)}]{Margalit2018b}
{Margalit}, B., \& {Metzger}, B.~D. 2018, \apj, 868, L4,
  \dodoi{10.3847/2041-8213/aaedad}

\bibitem[{{Margalit} {et~al.}(2018){Margalit}, {Metzger}, {Berger}, {Nicholl},
  {Eftekhari}, \& {Margutti}}]{Margalit2018a}
{Margalit}, B., {Metzger}, B.~D., {Berger}, E., {et~al.} 2018, \mnras, 481,
  2407, \dodoi{10.1093/mnras/sty2417}

\bibitem[{{Masui} {et~al.}(2015){Masui}, {Lin}, {Sievers}, {Anderson}, {Chang},
  {Chen}, {Ganguly}, {Jarvis}, {Kuo}, {Li}, {Liao}, {McLaughlin}, {Pen},
  {Peterson}, {Roman}, {Timbie}, {Voytek}, \& {Yadav}}]{Masui2015}
{Masui}, K., {Lin}, H.-H., {Sievers}, J., {et~al.} 2015, \nat, 528, 523,
  \dodoi{10.1038/nature15769}

\bibitem[{{Medin} \& {Lai}(2010)}]{Medin2010}
{Medin}, Z., \& {Lai}, D. 2010, \mnras, 406, 1379,
  \dodoi{10.1111/j.1365-2966.2010.16776.x}

\bibitem[{{M{\'e}sz{\'a}ros} \& {Rees}(1997)}]{Meszaros1997}
{M{\'e}sz{\'a}ros}, P., \& {Rees}, M.~J. 1997, \apj, 476, 232,
  \dodoi{10.1086/303625}

\bibitem[{{Metzger} {et~al.}(2017){Metzger}, {Berger}, \&
  {Margalit}}]{Metzger2017}
{Metzger}, B.~D., {Berger}, E., \& {Margalit}, B. 2017, \apj, 841, 14,
  \dodoi{10.3847/1538-4357/aa633d}

\bibitem[{{Metzger} {et~al.}(2019){Metzger}, {Margalit}, \&
  {Sironi}}]{Metzger2019}
{Metzger}, B.~D., {Margalit}, B., \& {Sironi}, L. 2019, arXiv e-prints.
\newblock \doarXiv{1902.01866}

\bibitem[{{Michilli} {et~al.}(2018){Michilli}, {Seymour}, {Hessels}, {Spitler},
  {Gajjar}, {Archibald}, {Bower}, {Chatterjee}, {Cordes}, {Gourdji}, {Heald},
  {Kaspi}, {Law}, {Sobey}, {Adams}, {Bassa}, {Bogdanov}, {Brinkman},
  {Demorest}, {Fernand ez}, {Hellbourg}, {Lazio}, {Lynch}, {Maddox}, {Marcote},
  {McLaughlin}, {Paragi}, {Ransom}, {Scholz}, {Siemion}, {Tendulkar}, {van
  Rooy}, {Wharton}, \& {Whitlow}}]{Michilli2018}
{Michilli}, D., {Seymour}, A., {Hessels}, J.~W.~T., {et~al.} 2018, \nat, 553,
  182, \dodoi{10.1038/nature25149}

\bibitem[{{Mottez} \& {Zarka}(2014)}]{Mottez2014}
{Mottez}, F., \& {Zarka}, P. 2014, Astronomy and Astrophysics, 569, A86,
  \dodoi{10.1051/0004-6361/201424104}

\bibitem[{{Murase} {et~al.}(2016){Murase}, {Kashiyama}, \&
  {M{\'e}sz{\'a}ros}}]{Murase2016}
{Murase}, K., {Kashiyama}, K., \& {M{\'e}sz{\'a}ros}, P. 2016, \mnras, 461,
  1498, \dodoi{10.1093/mnras/stw1328}

\bibitem[{{Nicholl} {et~al.}(2017){Nicholl}, {Williams}, {Berger}, {Villar},
  {Alexander}, {Eftekhari}, \& {Metzger}}]{Nicholl2017}
{Nicholl}, M., {Williams}, P.~K.~G., {Berger}, E., {et~al.} 2017, \apj, 843,
  84, \dodoi{10.3847/1538-4357/aa794d}

\bibitem[{{Petroff} {et~al.}(2016){Petroff}, {Barr}, {Jameson}, {Keane},
  {Bailes}, {Kramer}, {Morello}, {Tabbara}, \& {van Straten}}]{Petroff2016}
{Petroff}, E., {Barr}, E.~D., {Jameson}, A., {et~al.} 2016, \pasa, 33, e045,
  \dodoi{10.1017/pasa.2016.35}

\bibitem[{{Petroff} {et~al.}(2017){Petroff}, {Burke-Spolaor}, {Keane},
  {McLaughlin}, {Miller}, {Andreoni}, {Bailes}, {Barr}, {Bernard}, {Bhandari},
  {Bhat}, {Burgay}, {Caleb}, {Champion}, {Chandra}, {Cooke}, {Dhillon},
  {Farnes}, {Hardy}, {Jaroenjittichai}, {Johnston}, {Kasliwal}, {Kramer},
  {Littlefair}, {Macquart}, {Mickaliger}, {Possenti}, {Pritchard}, {Ravi},
  {Rest}, {Rowlinson}, {Sawangwit}, {Stappers}, {Sullivan}, {Tiburzi}, {van
  Straten}, {ANTARES Collaboration}, {Albert}, {Andr{\'e}}, {Anghinolfi},
  {Anton}, {Ardid}, {Aubert}, {Avgitas}, {Baret}, {Barrios-Mart{\'\i}}, {Basa},
  {Bertin}, {Biagi}, {Bormuth}, {Bourret}, {Bouwhuis}, {Bruijn}, {Brunner},
  {Busto}, {Capone}, {Caramete}, {Carr}, {Celli}, {Chiarusi}, {Circella},
  {Coelho}, {Coleiro}, {Coniglione}, {Costantini}, {Coyle}, {Creusot},
  {Deschamps}, {de Bonis}, {Distefano}, {di Palma}, {Donzaud}, {Dornic},
  {Drouhin}, {Eberl}, {El Bojaddaini}, {Els{\"a}sser}, {Enzenh{\"o}fer},
  {Felis}, {Fusco}, {Galat{\`a}}, {Gay}, {Gei{\ss}els{\"o}der}, {Geyer},
  {Giordano}, {Gleixner}, {Glotin}, {Gr{\'e}goire}, {Gracia-Ruiz}, {Graf},
  {Hallmann}, {van Haren}, {Heijboer}, {Hello}, {Hern{\'a}ndez-Rey},
  {H{\"o}{\ss}l}, {Hofest{\"a}dt}, {Hugon}, {Illuminati}, {James}, {de Jong},
  {Jongen}, {Kadler}, {Kalekin}, {Katz}, {Kie{\ss}ling}, {Kouchner}, {Kreter},
  {Kreykenbohm}, {Kulikovskiy}, {Lachaud}, {Lahmann}, {Lef{\`e}vre}, {Leonora},
  {Lotze}, {Loucatos}, {Marcelin}, {Margiotta}, {Marinelli},
  {Mart{\'\i}nez-Mora}, {Mathieu}, {Mele}, {Melis}, {Michael}, {Migliozzi},
  {Moussa}, {Mueller}, {Nezri}, {P{\v{a}}v{\v{a}}la{\c{s}}}, {Pellegrino},
  {Perrina}, {Piattelli}, {Popa}, {Pradier}, {Quinn}, {Racca}, {Riccobene},
  {Roensch}, {S{\'a}nchez-Losa}, {Salda{\~n}a}, {Salvadori}, {Samtleben},
  {Sanguineti}, {Sapienza}, {Schnabel}, {Seitz}, {Sieger}, {Spurio},
  {Stolarczyk}, {Taiuti}, {Tayalati}, {Trovato}, {Tselengidou}, {Turpin},
  {T{\"o}nnis}, {Vallage}, {Vall{\'e}e}, {van Elewyck}, {Vivolo}, {Vizzoca},
  {Wagner}, {Wilms}, {Zornoza}, {Z{\'u}{\~n}iga}, {H.~E.~S.~S. Collaboration},
  {Abdalla}, {Abramowski}, {Aharonian}, {Ait Benkhali}, {Akhperjanian},
  {Andersson}, {Ang{\"u}ner}, {Arrieta}, {Aubert}, {Backes}, {Balzer},
  {Barnard}, {Becherini}, {Tjus}, {Berge}, {Bernhard}, {Bernl{\"o}hr},
  {Blackwell}, {B{\"o}ttcher}, {Boisson}, {Bolmont}, {Bordas}, {Bregeon},
  {Brun}, {Brun}, {Bryan}, {Bulik}, {Capasso}, {Casanova}, {Cerruti},
  {Chakraborty}, {Chalme-Calvet}, {Chaves}, {Chen}, {Chevalier},
  {Chr{\'e}tien}, {Colafrancesco}, {Cologna}, {Condon}, {Conrad}, {Cui},
  {Davids}, {Decock}, {Degrange}, {Deil}, {Devin}, {Dewilt}, {Dirson},
  {Djannati-Ata{\"\i}}, {Domainko}, {Donath}, {Drury}, {Dubus}, {Dutson},
  {Dyks}, {Edwards}, {Egberts}, {Eger}, {Ernenwein}, {Eschbach}, {Farnier},
  {Fegan}, {Fernandes}, {Fiasson}, {Fontaine}, {F{\"o}rster}, {Funk},
  {F{\"u}{\ss}ling}, {Gabici}, {Gajdus}, {Gallant}, {Garrigoux}, {Giavitto},
  {Giebels}, {Glicenstein}, {Gottschall}, {Goyal}, {Grondin}, {Hadasch},
  {Hahn}, {Haupt}, {Hawkes}, {Heinzelmann}, {Henri}, {Hermann}, {Hervet},
  {Hinton}, {Hofmann}, {Hoischen}, {Holler}, {Horns}, {Ivascenko},
  {Jacholkowska}, {Jamrozy}, {Janiak}, {Jankowsky}, {Jankowsky}, {Jingo},
  {Jogler}, {Jouvin}, {Jung-Richardt}, {Kastendieck}, {Katarzy{\'n}ski},
  {Kerszberg}, {Kh{\'e}lifi}, {Kieffer}, {King}, {Klepser}, {Klochkov},
  {Klu{\'z}niak}, {Kolitzus}, {Komin}, {Kosack}, {Krakau}, {Kraus}, {Krayzel},
  {Kr{\"u}ger}, {Laffon}, {Lamanna}, {Lau}, {Lees}, {Lefaucheur}, {Lefranc},
  {Lemi{\`e}re}, {Lemoine-Goumard}, {Lenain}, {Leser}, {Lohse}, {Lorentz},
  {Liu}, {L{\'o}pez-Coto}, {Lypova}, {Marandon}, {Marcowith}, {Mariaud},
  {Marx}, {Maurin}, {Maxted}, {Mayer}, {Meintjes}, {Meyer}, {Mitchell},
  {Moderski}, {Mohamed}, {Mohrmann}, {Mor{\^a}}, {Moulin}, {Murach}, {de
  Naurois}, {Niederwanger}, {Niemiec}, {Oakes}, {O'Brien}, {Odaka}, {{\"O}ttl},
  {Ohm}, {Ostrowski}, {Oya}, {Padovani}, {Panter}, {Parsons}, {Pekeur},
  {Pelletier}, {Perennes}, {Petrucci}, {Peyaud}, {Piel}, {Pita}, {Poon},
  {Prokhorov}, {Prokoph}, {P{\"u}hlhofer}, {Punch}, {Quirrenbach}, {Raab},
  {Reimer}, {Reimer}, {Renaud}, {Reyes}, {Rieger}, {Romoli}, {Rosier-Lees},
  {Rowell}, {Rudak}, {Rulten}, {Sahakian}, {Salek}, {Sanchez}, {Santangelo},
  {Sasaki}, {Schlickeiser}, {Schulz}, {Sch{\"u}ssler}, {Schwanke}, {Schwemmer},
  {Settimo}, {Seyffert}, {Shafi}, {Shilon}, {Simoni}, {Sol}, {Spanier},
  {Spengler}, {Spies}, {Stawarz}, {Steenkamp}, {Stegmann}, {Stinzing}, {Stycz},
  {Sushch}, {Tavernet}, {Tavernier}, {Taylor}, {Terrier}, {Tibaldo}, {Tiziani},
  {Tluczykont}, {Trichard}, {Tuffs}, {Uchiyama}, {Walt}, {van Eldik}, {van
  Rensburg}, {van Soelen}, {Vasileiadis}, {Veh}, {Venter}, {Viana}, {Vincent},
  {Vink}, {Voisin}, {V{\"o}lk}, {Vuillaume}, {Wadiasingh}, {Wagner}, {Wagner},
  {Wagner}, {White}, {Wierzcholska}, {Willmann}, {W{\"o}rnlein}, {Wouters},
  {Yang}, {Zabalza}, {Zaborov}, {Zacharias}, {Zanin}, {Zdziarski}, {Zech},
  {Zefi}, {Ziegler}, \& {{\.Z}ywucka}}]{Petroff2017}
{Petroff}, E., {Burke-Spolaor}, S., {Keane}, E.~F., {et~al.} 2017, \mnras, 469,
  4465, \dodoi{10.1093/mnras/stx1098}

\bibitem[{{Platts} {et~al.}(2018){Platts}, {Weltman}, {Walters}, {Tendulkar},
  {Gordin}, \& {Kandhai}}]{Platts2018}
{Platts}, E., {Weltman}, A., {Walters}, A., {et~al.} 2018, arXiv e-prints,
  arXiv:1810.05836.
\newblock \doarXiv{1810.05836}

\bibitem[{{Plotnikov} \& {Sironi}(2019)}]{Plotnikov2019}
{Plotnikov}, I., \& {Sironi}, L. 2019, \mnras, \dodoi{10.1093/mnras/stz640}

\bibitem[{{Popov} \& {Postnov}(2013)}]{Popov2013}
{Popov}, S.~B., \& {Postnov}, K.~A. 2013, arXiv e-prints, arXiv:1307.4924.
\newblock \doarXiv{1307.4924}

\bibitem[{{Ravi} {et~al.}(2015){Ravi}, {Shannon}, \& {Jameson}}]{Ravi2015}
{Ravi}, V., {Shannon}, R.~M., \& {Jameson}, A. 2015, \apj, 799, L5,
  \dodoi{10.1088/2041-8205/799/1/L5}

\bibitem[{{Ravi} {et~al.}(2016){Ravi}, {Shannon}, {Bailes}, {Bannister},
  {Bhandari}, {Bhat}, {Burke-Spolaor}, {Caleb}, {Flynn}, {Jameson}, {Johnston},
  {Keane}, {Kerr}, {Tiburzi}, {Tuntsov}, \& {Vedantham}}]{Ravi2016}
{Ravi}, V., {Shannon}, R.~M., {Bailes}, M., {et~al.} 2016, Science, 354, 1249,
  \dodoi{10.1126/science.aaf6807}

\bibitem[{{Ravi} {et~al.}(2019){Ravi}, {Catha}, {D'Addario}, {Djorgovski},
  {Hallinan}, {Hobbs}, {Kocz}, {Kulkarni}, {Shi}, \& {Vedantham}}]{Ravi2019}
{Ravi}, V., {Catha}, M., {D'Addario}, L., {et~al.} 2019, arXiv e-prints,
  arXiv:1907.01542.
\newblock \doarXiv{1907.01542}

\bibitem[{{Rybicki} {et~al.}(1986){Rybicki}, {Lightman}, \&
  {Paul}}]{Rybicki1986}
{Rybicki}, G.~B., {Lightman}, A.~P., \& {Paul}, H.~G. 1986, Astronomische
  Nachrichten, 307, 170

\bibitem[{{Sari} {et~al.}(1998){Sari}, {Piran}, \& {Narayan}}]{Sari1998}
{Sari}, R., {Piran}, T., \& {Narayan}, R. 1998, \apjl, 497, L17,
  \dodoi{10.1086/311269}

\bibitem[{{Scholz} {et~al.}(2016){Scholz}, {Spitler}, {Hessels}, {Chatterjee},
  {Cordes}, {Kaspi}, {Wharton}, {Bassa}, {Bogdanov}, {Camilo}, {Crawford},
  {Deneva}, {van Leeuwen}, {Lynch}, {Madsen}, {McLaughlin}, {Mickaliger},
  {Parent}, {Patel}, {Ransom}, {Seymour}, {Stairs}, {Stappers}, \&
  {Tendulkar}}]{Scholz2016}
{Scholz}, P., {Spitler}, L.~G., {Hessels}, J.~W.~T., {et~al.} 2016, \apj, 833,
  177, \dodoi{10.3847/1538-4357/833/2/177}

\bibitem[{{Shannon} {et~al.}(2018){Shannon}, {Macquart}, {Bannister}, {Ekers},
  {James}, {Os{\l}owski}, {Qiu}, {Sammons}, {Hotan}, {Voronkov}, {Beresford},
  {Brothers}, {Brown}, {Bunton}, {Chippendale}, {Haskins}, {Leach},
  {Marquarding}, {McConnell}, {Pilawa}, {Sadler}, {Troup}, {Tuthill},
  {Whiting}, {Allison}, {Anderson}, {Bell}, {Collier}, {G{\"u}rkan}, {Heald},
  \& {Riseley}}]{Shannon2018}
{Shannon}, R.~M., {Macquart}, J.~P., {Bannister}, K.~W., {et~al.} 2018, \nat,
  562, 386, \dodoi{10.1038/s41586-018-0588-y}

\bibitem[{{Shende} {et~al.}(2019){Shende}, {Subramanian}, \&
  {Sachdeva}}]{Shende2019}
{Shende}, M.~B., {Subramanian}, P., \& {Sachdeva}, N. 2019, arXiv e-prints,
  arXiv:1904.10870.
\newblock \doarXiv{1904.10870}

\bibitem[{{Spitler} {et~al.}(2014){Spitler}, {Cordes}, {Hessels}, {Lorimer},
  {McLaughlin}, {Chatterjee}, {Crawford}, {Deneva}, {Kaspi}, {Wharton},
  {Allen}, {Bogdanov}, {Brazier}, {Camilo}, {Freire}, {Jenet},
  {Karako-Argaman}, {Knispel}, {Lazarus}, {Lee}, {van Leeuwen}, {Lynch},
  {Ransom}, {Scholz}, {Siemens}, {Stairs}, {Stovall}, {Swiggum},
  {Venkataraman}, {Zhu}, {Aulbert}, \& {Fehrmann}}]{Spitler2014}
{Spitler}, L.~G., {Cordes}, J.~M., {Hessels}, J.~W.~T., {et~al.} 2014, \apj,
  790, 101, \dodoi{10.1088/0004-637X/790/2/101}

\bibitem[{{Spitler} {et~al.}(2016){Spitler}, {Scholz}, {Hessels}, {Bogdanov},
  {Brazier}, {Camilo}, {Chatterjee}, {Cordes}, {Crawford}, {Deneva}, {Ferdman},
  {Freire}, {Kaspi}, {Lazarus}, {Lynch}, {Madsen}, {McLaughlin}, {Patel},
  {Ransom}, {Seymour}, {Stairs}, {Stappers}, {van Leeuwen}, \&
  {Zhu}}]{Spitler2016}
{Spitler}, L.~G., {Scholz}, P., {Hessels}, J.~W.~T., {et~al.} 2016, \nat, 531,
  202, \dodoi{10.1038/nature17168}

\bibitem[{{Tendulkar} {et~al.}(2017){Tendulkar}, {Bassa}, {Cordes}, {Bower},
  {Law}, {Chatterjee}, {Adams}, {Bogdanov}, {Burke-Spolaor}, {Butler},
  {Demorest}, {Hessels}, {Kaspi}, {Lazio}, {Maddox}, {Marcote}, {McLaughlin},
  {Paragi}, {Ransom}, {Scholz}, {Seymour}, {Spitler}, {van Langevelde}, \&
  {Wharton}}]{Tendulkar2017}
{Tendulkar}, S.~P., {Bassa}, C.~G., {Cordes}, J.~M., {et~al.} 2017, \apj, 834,
  L7, \dodoi{10.3847/2041-8213/834/2/L7}

\bibitem[{{Thornton} {et~al.}(2013){Thornton}, {Stappers}, {Bailes},
  {Barsdell}, {Bates}, {Bhat}, {Burgay}, {Burke-Spolaor}, {Champion}, {Coster},
  {D'Amico}, {Jameson}, {Johnston}, {Keith}, {Kramer}, {Levin}, {Milia}, {Ng},
  {Possenti}, \& {van Straten}}]{Thornton2013}
{Thornton}, D., {Stappers}, B., {Bailes}, M., {et~al.} 2013, Science, 341, 53,
  \dodoi{10.1126/science.1236789}

\bibitem[{{Wang} {et~al.}(2018){Wang}, {Peng}, {Wu}, \& {Dai}}]{Wang2018}
{Wang}, J.-S., {Peng}, F.-K., {Wu}, K., \& {Dai}, Z.-G. 2018, \apj, 868, 19,
  \dodoi{10.3847/1538-4357/aae531}

\bibitem[{{Wang} {et~al.}(2016){Wang}, {Yang}, {Wu}, {Dai}, \&
  {Wang}}]{Wang2016}
{Wang}, J.-S., {Yang}, Y.-P., {Wu}, X.-F., {Dai}, Z.-G., \& {Wang}, F.-Y. 2016,
  \apj, 822, L7, \dodoi{10.3847/2041-8205/822/1/L7}

\bibitem[{{Waxman}(2017)}]{Waxman2017}
{Waxman}, E. 2017, \apj, 842, 34, \dodoi{10.3847/1538-4357/aa713e}

\bibitem[{{Wu} {et~al.}(2003){Wu}, {Dai}, {Huang}, \& {Lu}}]{Wu2003}
{Wu}, X.~F., {Dai}, Z.~G., {Huang}, Y.~F., \& {Lu}, T. 2003, \mnras, 342, 1131,
  \dodoi{10.1046/j.1365-8711.2003.06602.x}

\bibitem[{{Xiao} \& {Dai}(2017)}]{Xiao2017}
{Xiao}, D., \& {Dai}, Z.-G. 2017, \apj, 846, 130,
  \dodoi{10.3847/1538-4357/aa8625}

\bibitem[{{Yang} \& {Dai}(2019)}]{YangDai2019}
{Yang}, Y.-H., \& {Dai}, Z.-G. 2019, \apj, 885, 149,
  \dodoi{10.3847/1538-4357/ab48dd}

\bibitem[{{Yang} \& {Zhang}(2018)}]{Yang2018}
{Yang}, Y.-P., \& {Zhang}, B. 2018, \apj, 868, 31,
  \dodoi{10.3847/1538-4357/aae685}

\bibitem[{{Yang} {et~al.}(2016){Yang}, {Zhang}, \& {Dai}}]{Yang2016}
{Yang}, Y.-P., {Zhang}, B., \& {Dai}, Z.-G. 2016, \apjl, 819, L12,
  \dodoi{10.3847/2041-8205/819/1/L12}

\bibitem[{{Yang} {et~al.}(2019){Yang}, {Zhang}, \& {Wei}}]{Yang2019}
{Yang}, Y.-P., {Zhang}, B., \& {Wei}, J.-Y. 2019, \apj, 878, 89,
  \dodoi{10.3847/1538-4357/ab1fe2}

\bibitem[{{Zhang}(2016)}]{Zhang2016}
{Zhang}, B. 2016, \apj, 827, L31, \dodoi{10.3847/2041-8205/827/2/L31}

\bibitem[{{Zhang}(2017)}]{Zhang2017}
---. 2017, \apj, 836, L32, \dodoi{10.3847/2041-8213/aa5ded}

\bibitem[{{Zhang} {et~al.}(2019){Zhang}, {Hobbs}, {Dai}, {Toomey},
  {Staveley-Smith}, {Russell}, \& {Wu}}]{Zhang2019}
{Zhang}, S.~B., {Hobbs}, G., {Dai}, S., {et~al.} 2019, \mnras, 484, L147,
  \dodoi{10.1093/mnrasl/slz023}

\bibitem[{{Zhang} {et~al.}(2008){Zhang}, {Woosley}, \& {Heger}}]{Zhang2008}
{Zhang}, W., {Woosley}, S.~E., \& {Heger}, A. 2008, \apj, 679, 639,
  \dodoi{10.1086/526404}

\end{thebibliography}

\end{document}